\documentclass[number,sort&compress,5p,twocolumn,times]{elsarticle}
\usepackage[binary-units=true,separate-uncertainty=true]{siunitx}
\usepackage{lineno,hyperref}
\usepackage{tikz}
\usepackage{amssymb,amsmath,mathtools}
\usepackage{caption}
\usepackage{xspace}
\usepackage{booktabs}
\usepackage{siunitx}
\usepackage{textcomp}
\usepackage[percent]{overpic}
\usepackage[list=true,listformat=simple,margin=10pt]{subcaption}
\usepackage{hepnames}

\journal{Nucl. Instr. Meth. A}

\begin{document}
\begin{frontmatter}
  \title{Technologies for Future Vertex and Tracking Detectors at CLIC}

\author{S.~Spannagel}
\ead{simon.spannagel@cern.ch}
\author{on behalf of the CLICdp Collaboration}

\address{CERN, Geneva, Switzerland}

\cortext[corr]{Corresponding author}

\begin{abstract}
  CLIC is a proposed linear $\Ppositron\Pelectron$ collider with center-of-mass energies of up to \SI{3}{\TeV}. Its main objectives are precise top quark and Higgs boson measurements, as well as searches for Beyond Standard Model physics. To meet the physics goals, the vertex and tracking detectors require not only a spatial resolution of a few micrometers and a very low material budget, but also timing capabilities with a precision of a few nanoseconds to allow suppression of beam-induced backgrounds.

  Different technologies using hybrid silicon detectors are explored for the vertex detectors, such as dedicated readout ASICs, small-pitch active edge sensors as well as capacitively coupled High-Voltage CMOS sensors. Monolithic sensors are considered as an option for the tracking detector, and a prototype using a CMOS process with a high-resistivity epitaxial layer is being designed. Different designs using a silicon-on-insulator process are under investigation for both vertex and tracking detector.

  All prototypes are evaluated in laboratory and beam tests, and newly developed simulation tools combining Geant4 and TCAD are used to assess and optimize their performance. This contribution gives an overview of the R\&D program for the CLIC vertex and tracking detectors, highlighting new results from the prototypes.
\end{abstract}


\end{frontmatter}

%% ==============================
\section{Introduction}
\label{sec:introduction}
%% ==============================

The Compact Linear Collider (CLIC)~\cite{clic,clic-baseline} is a proposed linear $\Ppositron\Pelectron$ collider to be built at CERN, Geneva, Switzerland.
It uses a novel two-beam acceleration scheme achieving field gradients of more than \SI{100}{\mega \volt \per \meter}, which enables the design of a compact accelerator.
The construction is foreseen in three stages with center-of-mass energies of \SI{380}{\GeV}, \SI{1.5}{\TeV} and \SI{3}{\TeV}, each with a different physics focus ranging from precision measurements of the Higgs boson and Top quark to exotica and beyond the Standard Model physics.

The design of the vertex and tracking detectors is strongly influenced by the beam structure at CLIC.
Trains of 312 bunches with a bunch spacing of \SI{0.5}{\nano \second} are repeated with a frequency of \SI{50}{\hertz}.
To obtain luminosities of up to \SI{5.9e34}{\per \centi\meter\squared\per\second}, the bunches are focused to a size of a few nanometers at the interaction point.
The resulting high bunch density leads to interactions between colliding bunches which create background particles.
At the highest collision energy, about 100 particles per bunch crossing are expected within the acceptance of the tracking systems, leading to pixel occupancies of up to 3\% per bunch train in the innermost layers.
Timing information is required to reduce the effect of these pile-up hits on the reconstruction.

The current model for a detector at CLIC~\cite{clicdet} features an all-silicon vertex and tracking system with very stringent requirements on resolution and material budget.
To limit multiple scattering, the material budget is limited to about 0.2\%\,X$_0$ per vertex detector layer and 1\%\,X$_0$ per layer of the tracking detector.
A low power consumption of below \SI{50}{\milli \watt \per \centi \meter \squared} in the vertex detector is required in order to enable passive air-flow cooling.

Single-point resolutions of about \SI{3}{\micro \meter} in the vertex and \SI{7}{\micro \meter} in the tracking detector are required for optimal flavor tagging and track reconstruction performance.
To allow efficient reduction of background, time stamping capabilities with a precision of about \SI{5}{\nano \second} are required.

The CLIC vertex and tracking detectors thus need to achieve an order of magnitude reduction both in material budget and cell size, while maintaining a similar hit-time resolution, compared to the current LHC detectors.

%% ==============================
\section{Silicon Technologies}
\label{sec:silicon}
%% ==============================

A variety of silicon technologies are currently explored in order to find the best match with the requirements for a detector at CLIC.
This section provides a brief overview of the considered technologies and showcases the investigated prototypes.

%% ==============================
\subsection{Hybrid Pixel Detectors}
\label{sec:hybrid}
%% ==============================

Hybrid detectors represent the traditional design of silicon pixel detectors in high-energy physics and consist of two independent parts: the high-resistivity sensor and the CMOS readout chip.
The connection between the individual channels is established by solder bumps.
With this design, extensive functionality can be placed in the pixels of the readout chip, while the sensor can be fully depleted to maximize charge collection.

Pixel cell sizes of \SI{25}{\micro \meter} -- \SI{250}{\micro \meter} have been achieved, but the bump bonding process is both the main cost driver and the limiting factor for the pixel pitch and the device thickness.

The \emph{CLICpix2} prototype~\cite{clicpix2} is a readout ASIC designed to meet the requirements for the CLIC vertex detector.
It features a $128 \times 128$ pixel matrix with a total active area of \SI{3.2 x 3.2}{\milli \meter}, implemented in a \SI{65}{\nano \meter} CMOS process with a pixel pitch of \SI{25}{\micro \meter} x \SI{25}{\micro \meter}.
The pixel cells feature simultaneous measurement of the charge and time-of-arrival.
The data acquisition is shutter-based, and the pixel matrix can be powered down between shutters.

\begin{figure}[tbp]
  \centering
  \resizebox{\columnwidth}{!}{
  \begin{overpic}[width=\columnwidth]{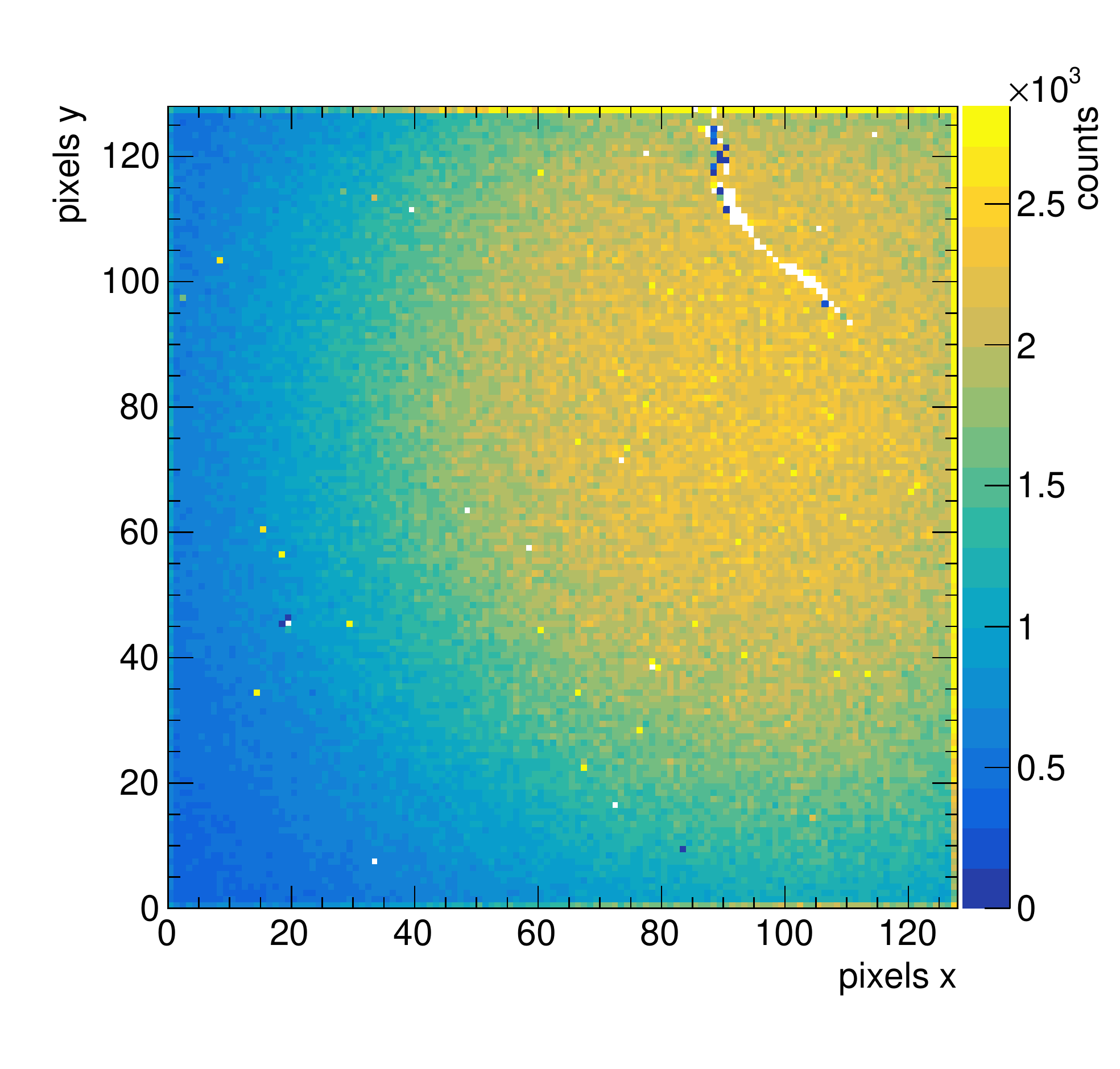}
      \put (20,23) {\textbf{CLICdp}}
      \put (20,20) {work in progress}
  \end{overpic}
  }
  \caption{Sr-90 measurement of a CLICpix2 prototype bonded to a planar sensor with \SI{130}{\micro \meter} thickness. A notch can be observed in the top-right corner, where 55 bumps were lost when lifting the chip from the tape. The isolated unresponsive pixels have been identified as dead channels of the readout ASIC.}
  \label{fig:clicpix2-source}
\end{figure}

Promising results have been achieved recently for single-chip bump bonding using a support wafer process with SnAg bumps, which is a challenge at the pixel pitch of \SI{25}{\micro \meter}, due to the requirements on the diameter and uniformity of the solder balls and on the alignment precision.
Figure~\ref{fig:clicpix2-source} shows a Sr-90 source measurement of a \emph{CLICpix2} prototype bonded to a planar silicon sensor, indicating a high bump yield of more than \SI{99.6}{\percent}.
The prototype has been successfully tested in laboratory and beam tests, its detailed characterization is still ongoing.

%% ==============================
\subsection{Monolithic High-Voltage CMOS Sensors}
%% ==============================

In monolithic High-Voltage (HV) CMOS sensors, electronics and sensor are placed on the same wafer~\cite{peric-hv-cmos}.
This facilitates fully integrated designs comprising amplification as well as discrimination and readout, which feature lower mass and do not require bump bonding.
The readout electronics is surrounded by a deep collection diode which also acts as shield from the electric field in the sensor bulk.
By this, higher bias voltages of about \SI{60}{\volt} can be applied to increase the depleted volume.
The large collection diode however entails a large input capacitance.
%, and full depletion has yet to be achieved.

\begin{figure}[tbp]
  \centering
  \resizebox{\columnwidth}{!}{
  \begin{overpic}[width=\columnwidth]{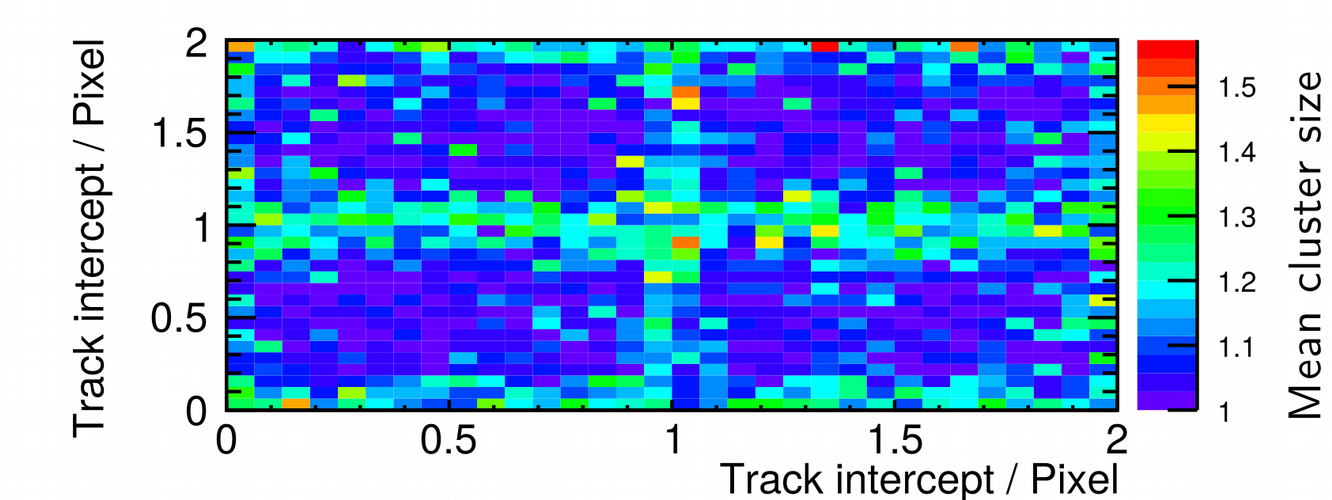}
      \put (20,13) {\textcolor{black!10}{\textbf{CLICdp}}}
      \put (20,10) {\textcolor{black!10}{work in progress}}
  \end{overpic}
  }
  \caption{Mean cluster size as a function of the incidence position of the particle for the \emph{ATLASpix\_Simple} prototype.
  Here, tracks from the full pixel matrix are plotted with respect to their relative position within a $2 \times 2$ pixel area.}
  \label{fig:atlaspix-clustersize}
\end{figure}

For the CLIC tracking detector, the \emph{ATLASpix\_Simple} prototype~\cite{peric-atlaspix} is being investigated.
This fully integrated chip designed for the ATLAS ITk upgrade has been produced in an AMS \SI{180}{\nano \meter} HV-CMOS process with substrate resistivities of $20 - \SI{1000}{\ohm \centi \meter}$.
The chip has 25 x 400 pixels with a pitch of \SI{130}{\micro \meter} x \SI{40}{\micro \meter} and features in-pixel charge amplifier and discriminators, while the charge and arrival time measurement is performed in the periphery.
Promising results have been achieved in first beam tests.
Figure~\ref{fig:atlaspix-clustersize} shows the mean cluster size as a function of the particle position within a $2 \times 2$ pixel area.
The expected charge sharing is clearly visible at the boundaries between pixel cells, where the mean cluster size increases while staying close to one inside the individual pixels.
A position resolution of \SI{13}{\micro \meter} has been reached, the efficiency of \SI{99.5}{\percent} has been measured at a charge threshold of \SI{1500}{e}.

The prototype is currently being integrated into the CaRIBou DAQ system~\cite{caribou} which allows to characterize the timing performance and provides the capacity to record data at higher rates.

%% ==============================
\subsection{Monolithic High-Resistivity CMOS Sensors}
%% ==============================

Monolithic High-Resistivity (HR) CMOS sensors represent an alternative to HV-CMOS sensors.
Here, the electronics is placed outside the charge-collection diode and is separately shielded.
Depletion is achieved by using a high-resistivity substrate and a lower bias voltage, and no dedicated high-voltage design rules by the foundry have to be taken into account.
The small collection diode reduces the input capacitance and thus reduces the noise.
Recent process modifications allow full lateral depletion~\cite{tj-modified}.

The \emph{Investigator} chip designed by the ALICE collaboration has been used to evaluate this technology in view of the CLIC tracking detector.
The chip is an analog prototype, where the digitization of the signals is performed off-chip in the data acquisition system.
Two variants of a TowerJazz \SI{180}{\nano \meter} HR-CMOS process have been investigated and resolutions of \SI{4}{\micro \meter} in space and \SI{5}{\nano \second} in time have been measured for a pitch of \SI{28 x 28}{\micro \meter}~\cite{CLICdp-Note-2017-004}.

A prototype of a fully integrated sensor dedicated to the CLIC tracking detector, \emph{CLICTD},  is currently being designed in this technology exploiting the very low noise and minimum detection threshold.
Given the excellent achievable resolution, this technology is also of interest for the CLIC vertex detector.

%% ==============================
\subsection{Monolithic Silicon-on-Insulator Sensors}
%% ==============================

Another option to isolate the sensitive CMOS electronics from the high-field region in the sensor is the silicon-on-insulator (SOI) approach, where the electronics is separated by an insulation oxide layer from the sensor on a single high-resistivity wafer.

\begin{figure}[tbp]
  \centering
  \resizebox{\columnwidth}{!}{
  \begin{overpic}[width=\columnwidth]{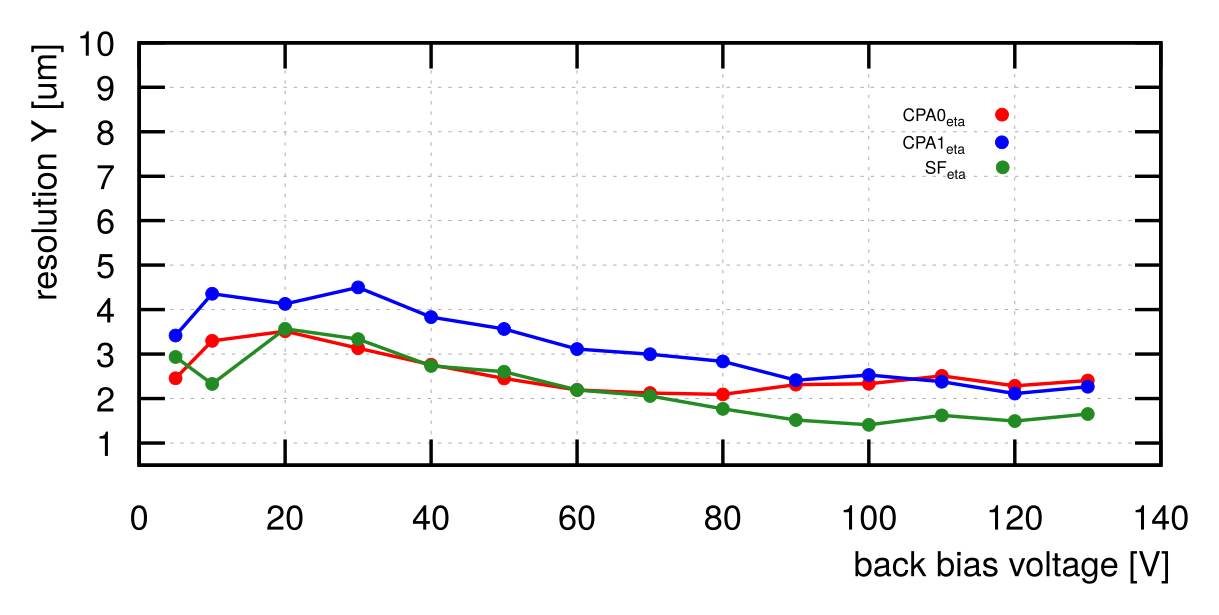}
      \put (16,40) {\textbf{CLICdp}}
      \put (16,37) {work in progress}
  \end{overpic}
  }
  \caption{Position resolution of the SOI test chip as a function of the applied back bias voltage. The three data sets correspond to the different pixel circuit designs, two with a charge pre-amplifier (CPA) and one source follower (SF) design.}
  \label{fig:soi-resolution}
\end{figure}

To evaluate this technology, an SOI test chip has been manufactured in a LAPIS \SI{200}{\nano \meter} SOI process~\cite{soi1}.
It combines different parameters on a single chip, such as different pixel pitches as well as different amplification and readout schemes.
First test beam results for a sensor with a thickness of \SI{500}{\micro \meter} and a pitch of \SI{30 x 30}{\micro \meter} show an excellent position resolution of about \SI{2}{\micro \meter} as indicated in Figure~\ref{fig:soi-resolution}.

Analysis of prototype test beam data is ongoing, and a dedicated prototype chip for the CLIC vertex detector, \emph{CLIPS}, has been designed and is currently in production.

%% ==============================
\subsection{Capacitively Coupled Detectors}
%% ==============================

In order to overcome the limitations of bump bonding in traditional hybrid detectors, capacitively coupled hybrids are being investigated, where an active HV-CMOS sensor with analog signal amplification is glued to a CMOS readout chip and the amplified signal coupled capacitively through the glue layer.

This approach permits using the full feature set of standard CMOS processes for the readout chip while keeping the required circuitry in the HV-CMOS chip to a minimum.
Investigations have shown that precise alignment and uniform glue dispensation are key to a good assembly performance.

\begin{figure}[tbp]
  \centering
  \resizebox{\columnwidth}{!}{
  \begin{overpic}[width=\columnwidth]{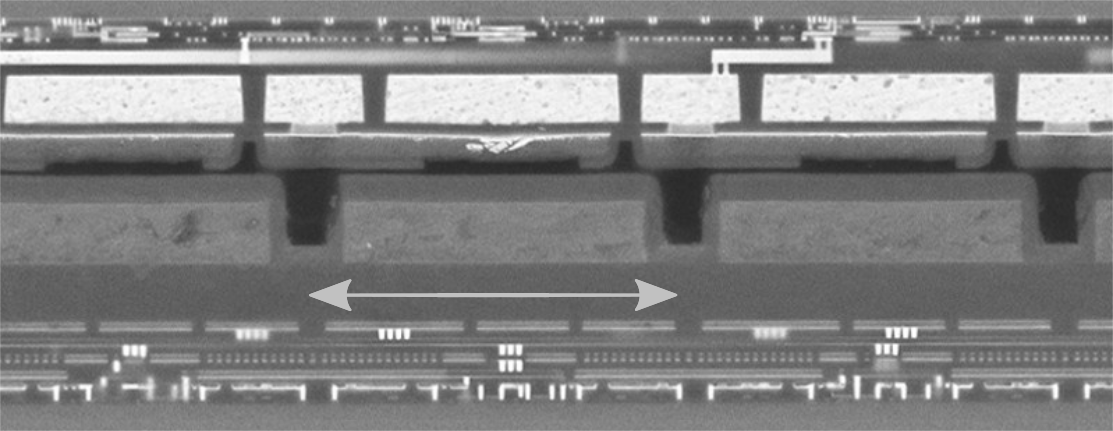}
      \put (75,30) {\colorbox{black!10}{CLICpix2}}
      \put (75,15) {\colorbox{black!10}{C3PD}}
      \put (38,15) {\textcolor{black!10}{\SI{25}{\micro \meter}}}
  \end{overpic}
  }
  \caption{Cross-section of a glue assembly with \emph{CLICpix2} and \emph{C3PD} with a pixel pitch of \SI{25}{\micro \meter}.}
  \label{fig:c3pd-glued}
\end{figure}

\emph{CLICpix2} prototypes have been glued to \emph{C3PD} sensors~\cite{c3pd}, a second-generation active sensor designed in an AMS \SI{180}{\nano \meter} HV-CMOS process with a matching pixel pitch of \SI{25 x 25}{\micro \meter}.
A cross-section of such an assembly is shown in Figure~\ref{fig:c3pd-glued}.
First test beam measurements show a temporal and spatial resolution of approximately \SI{7}{\nano \second} and \SI{8}{\micro \meter}, respectively.
Extensive simulations of the sensor using TCAD~\cite{buckland} and of the capacitive coupling via finite-element analysis~\cite{fe-sim} have been conducted for comparison with the measured performance.

Further assemblies with high-resistivity substrate sensors are under investigation, and the gluing process is being optimized further.

%% ==============================
\subsection{Enhanced Lateral Drift Detectors}
%% ==============================

A novel sensor concept explored for the CLIC vertex detector are Enhanced Lateral Drift (ELAD) sensors~\cite{elad-patent}, which are an attempt to break the paradigm of ever-decreasing pixel pitch.
In thin sensors, the position resolution is limited to $p / \sqrt{12}$, where $p$ is the pixel pitch, since almost no charge sharing takes place.

By introducing additional deep implants in the sensor bulk, the electric field is altered such that the lateral spread of charge carriers is increased during their drift.
The implants are designed in a way that the resulting charge sharing is close to the theoretical optimum of a linear charge distribution.
Challenges to overcome for this type of sensor are the complex production process and the exact placement of the implants to avoid low-field regions and the resulting recombination of charge carriers.

This sensor design performs very well in finite-element device simulation, and a first production batch with test structures as well as sensors with \SI{55}{\micro \meter} pitch is expected still in 2018.

%% ==============================
\section{Simulation Tools}
\label{sec:simulations}
%% ==============================

Advanced simulation tools help in understanding the performance of new prototypes and to optimize the design of new detectors.
The Monte-Carlo approach is widely used to account for the stochastic nature of the processes under investigation, and a new software framework has been developed to combine this with information obtained from detailed device modeling.

The Allpix Squared silicon detector simulation framework~\cite{allpix-squared} combines the detailed description of particle interactions provided by Geant4~\cite{geant4,geant4-2,geant4-3} with electric fields simulated using TCAD, and implements fast algorithms for charge carrier propagation in silicon as well as front-end electronics simulations.
It provides access to the main detector characteristics such as position resolution, charge collection efficiency, or tracking efficiency.

\begin{figure}[tbp]
  \centering
  \includegraphics[width=\columnwidth]{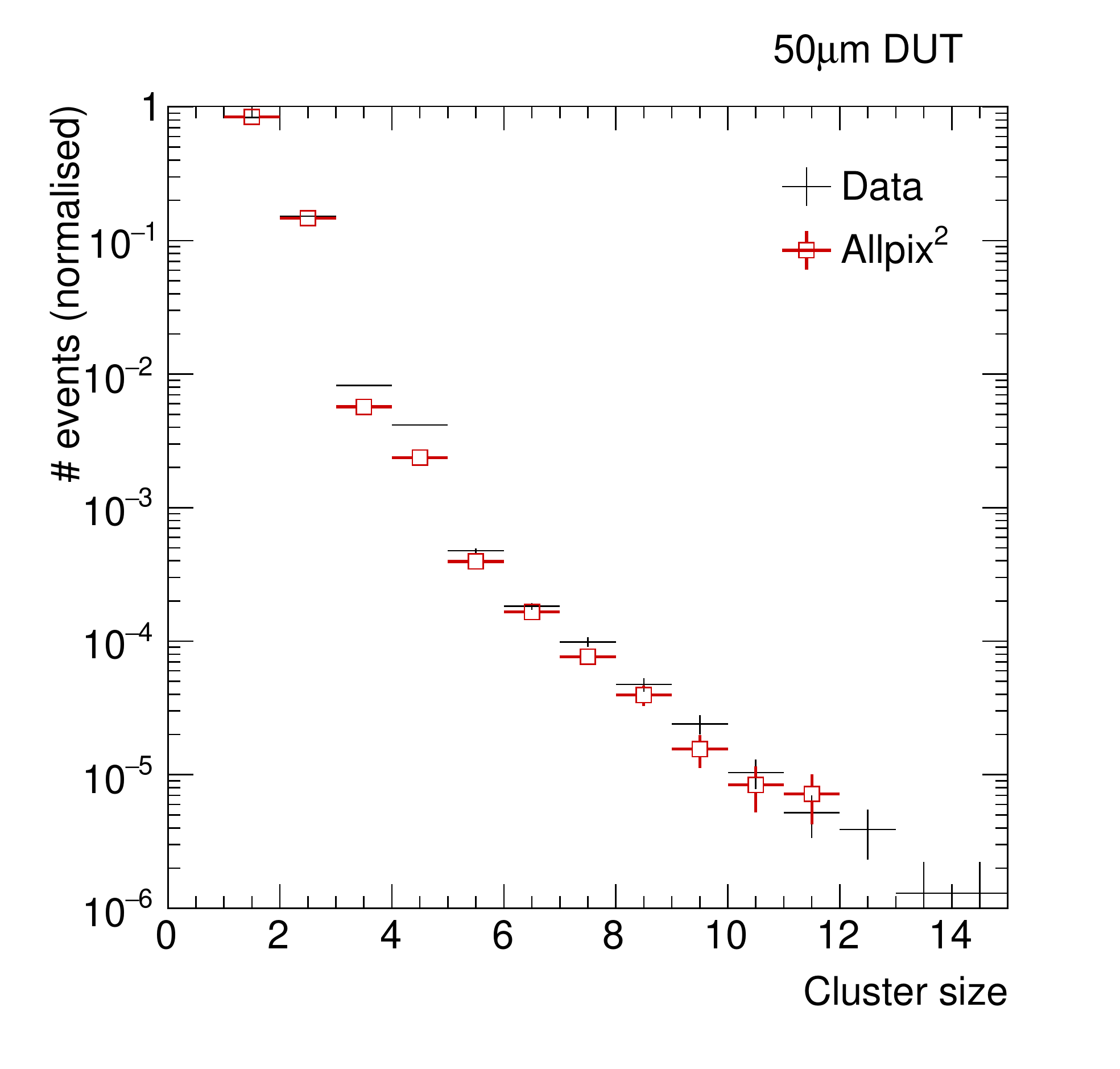}
  \caption{Comparison of the cluster size distribution from test beam measurements with a \SI{50}{\micro \meter} thick planar silicon sensor and a simulation of the device using Allpix Squared. Figure taken from~\cite{allpix-squared}.}
  \label{fig:allpix-validation}
\end{figure}

The simulation framework has been validated using test beam data, and very good agreement between data and simulation has been found.
An example is shown in Figure~\ref{fig:allpix-validation}, comparing the cluster size distribution of a \SI{50}{\micro \meter} thick planar silicon sensor as simulated and measured in test beam data.

The software sees continuous development and a new feature version has recently been released~\cite{allpix-1.2}.
Among other improvements, it adds support of magnetic fields and Lorentz drift, new particle source types and the possibility to record particle tracks as Monte Carlo truth information.

Currently, a detailed transient simulation using weighting fields is under development with the goal to better model the timing behavior of silicon detectors.

%% ==============================
\section{Summary and Outlook}
\label{sec:conclusions}
%% ==============================

The beam conditions and physics goals at the proposed CLIC linear $\Ppositron\Pelectron$ collider pose challenges to the vertex and tracking detectors.
An excellent spatial and temporal resolution is required while maintaining a minimum material budget and power consumption.

A comprehensive R\&D program is underway to qualify silicon technologies for the vertex and tracking detectors at CLIC.
Many different concepts are being investigated and most initial requirements have been shown to be achievable.
However, the desired position resolution of \SI{3}{\micro \meter} in the vertex detector remains a challenge, and prototypes fulfilling the full set of requirements are yet to be built.

A dedicated prototype using the HR-CMOS technology is currently being designed.
Productions of the SOI technology prototype as well as the first ELAD silicon sensor are ongoing.

New and validated simulation tools combining precise device modeling and the Monte-Carlo approach ease the R\&D and prototyping processes.

%% ==============================
\section{Acknowledgements}
\label{sec:acknowledgements}
%% ==============================

This project has received funding from the European Union's Horizon 2020 Research and Innovation programme under Grant Agreement no. 654168.

\section*{References}
\bibliography{bibliography}

\end{document}